\documentclass[12pt]{article}
\usepackage{graphicx}
\usepackage{braket}
\usepackage{slashed}
\usepackage{subfigure}
\usepackage{amsmath,amssymb,amsfonts}
\newcommand\pubnumber{UWThPh-2023-31}
\newcommand\pubdate{\today}

\def\institute{$^a$University of Vienna, Faculty of Physics, Boltzmanngasse 5, A-1090 Wien, Austria\\
$^b$University of Vienna, Erwin Schr\"odinger International Institute for Mathematics and Physics, Boltzmanngasse 9, A-1090 Wien, Austria \\
$^c$University of Graz, Faculty of Physics, NAWI Graz, Universit\"atsplatz 5, A-8010 Graz, Austria\\
$^d$University of Vienna, Vienna Doctoral School in Physics, Boltzmanngasse, 5 A-1090 Wien, Austria\\ }

\def\Title#1{\begin{center} {\Large #1 } \end{center}}
\def\Author#1{\begin{center}{ \sc #1} \end{center}}
\def\Address#1{\begin{center}{ \it #1} \end{center}}

\newcommand\pubblock{\rightline{\begin{tabular}{l} \pubnumber\\
         \pubdate  \end{tabular}}}
\newenvironment{Abstract}{\begin{quotation}  }{\end{quotation}}
\newenvironment{Presented}{\begin{quotation} \begin{center} 
             PRESENTED AT\end{center}\bigskip 
      \begin{center}\begin{large}}{\end{large}\end{center} \end{quotation}}
\def\Acknowledgements{\bigskip  \bigskip \begin{center} \begin{large}
             \bf ACKNOWLEDGEMENTS \end{large}\end{center}}
\def\beq{\begin{equation}}
\def\eeq#1{\label{#1}\end{equation}}
\def\eeqn{\end{equation}}

\def\beqa{\begin{eqnarray}}
\def\eeqa#1{\label{#1}\end{eqnarray}}
\def\eeqan{\end{eqnarray}}

\let\bar=\overbar

\def\bra#1{\left\langle{ #1} \right|}
\def\ket#1{\left| {#1} \right\rangle}

\def\Dslash{\not{\hbox{\kern-4pt $D$}}}
\def\dslash{\not{\hbox{\kern-2pt $\del$}}}

\def\msb{{\bar{\ssstyle M \kern -1pt S}}}

\newcommand{\MadGraph}{\textsc{MadGraph5}}
\begin{document}
\begin{titlepage}
\pubblock

\vfill
\Title{Beyond the Narrow-Width Limit for Off-Shell and Boosted Differential Top Quark Decays}
\vfill
%\Author{ExpampleAuthor\authemail}
\Author{ Andr\'e H. Hoang\footnote{andre.hoang@univie.ac.at}$^{,a, b}$, Simon Pl\"atzer\footnote{simon.plaetzer@uni-graz.at}$^{,a, c}$, Christoph Regner\footnote{christoph.regner@univie.ac.at}$^{,a, d}$, \\ Ines Ruffa\footnote{ines.ruffa@univie.ac.at}$^{,a, d}$}
\Address{\institute}
\vfill
\begin{Abstract}
The standard approaches for describing top quark production and its decay dynamics are currently mostly either based on the narrow-width (NW) limit or on off-shell fixed-order calculations. In this article we present a factorised approach for boosted top quarks that combines the properties of the NW limit and off-shell computations accounting for the dominant off-shell effects in an expansion in $m_t/Q$ with the hard scattering scale $Q$. We discuss the key ideas of our approach and show some preliminary results at tree-level.   
\end{Abstract}
\vfill
\begin{Presented}
$16^\mathrm{th}$ International Workshop on Top Quark Physics\\
(Top2023), 24--29 September, 2023
\end{Presented}
\vfill
\end{titlepage}
\def\thefootnote{\fnsymbol{footnote}}
\setcounter{footnote}{0}

\section{Introduction}
Due to its large mass the top quark plays an important role in consistency checks of the Standard Model as well as in new-physics searches. Studies concerning precise theoretical predictions of top production and its decay are commonly based on two different approaches, the narrow-width (NW) limit of the top quark propagator and full off-shell computations in fixed-order QCD.\\
The NW limit (see e.g.~\cite{Uhlemann:2008pm,Fuchs:2014ola}), in which the top quark is treated as an on-shell particle, allows for a convenient factorisation of the production and the decay dynamics albeit with the possibility to account for spin-correlations. It is used for current experimental analyses. However, it is known to be inadequate at kinematic endpoints and thresholds that are sensitive to the precise value of the invariant mass of the top-like intermediate state. This may be also relevant for the direct top mass measurements that have recently reached uncertainties below $400$ MeV~\cite{CMS:2022kqg}, which is much smaller than the top width $\Gamma_t \approx 1.4 \,\mathrm{GeV}$. \\
The full off-shell fixed-order calculations, on the other hand, take into account both resonant and non-resonant effects as well as non-factorizable contributions in their entirety. As a result, the computations are more involved and currently restricted to NLO QCD corrections, and not widely used in experimental analysis.\\
In this presentation we introduce a novel approach for boosted top quarks that allows to combine the factorization property of the NW limit and the dominant off-shell effects in an expansion in $m_t/Q$, where $Q$ is the hard scattering scale. The approach is valid in the resonance region for boosted top production and applies methodology known from soft-collinear effective theory (SCET) to the electroweak theory~\cite{Chiu:2007dg}. Eventually, the approach also allows to incorporate resummed QCD corrections for differential top decay observables, and it can also be combined with the spin-density formalism giving access to spin physics for off-shell top quarks. In the following, we discuss the basic ideas of our approach and present some results for simple observables at tree-level.
%%%%%%%%%%%%%%%%%%%%%%%%%%%%%%%%%%%%%%%%%%%%%%%%%%%%%%%%%%%%%%%%%%%%%%%%%
%%%%%%%%%%%%%%%%%%%%%%%%%%%%%%%%%%%%%%%%%%%%%%%%%%%%%%%%%%%%%%%%%%%%%%%%%%%
%%%%%%%%%%%%%%%%%%%%%%%%%%%%%%%%%%%%%%%%%%%%%%%%%%%%%%%%%%%%%%%%%%%%%%%%%%%
\section{Differential electro-weak top jet functions}

\begin{figure}
\centering
\includegraphics[height=3.5cm]{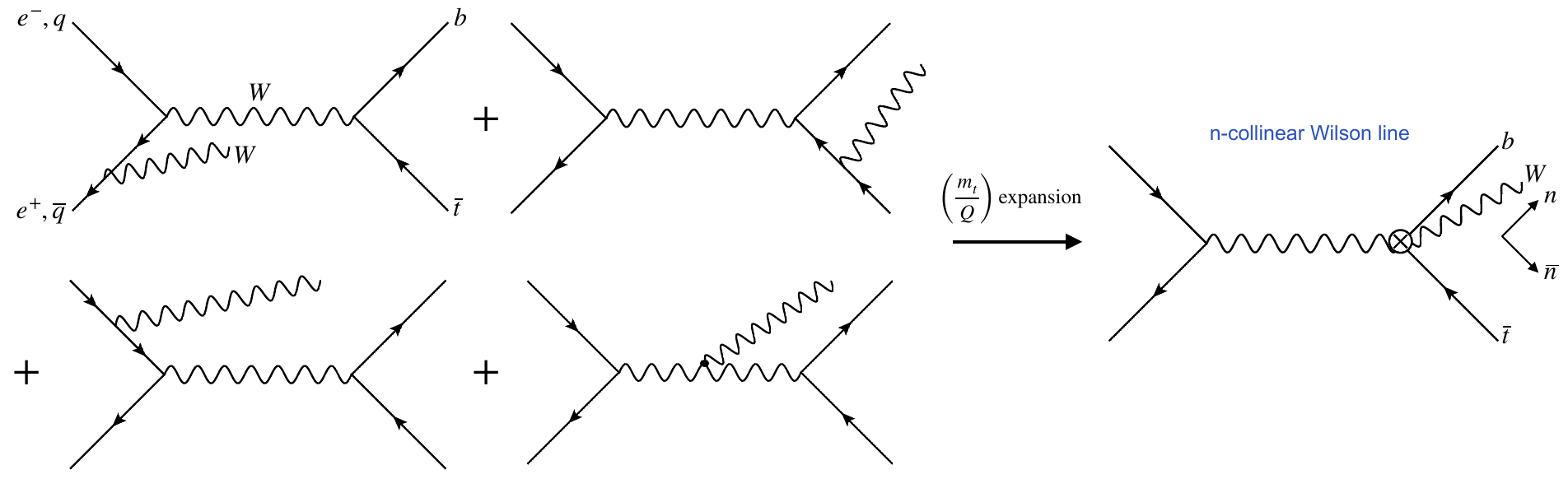}
\caption{Emergence of $n$-collinear Wilson line from a $2\rightarrow 2$ scattering with the emission of a $W$-boson collinear in top direction in an expansion $m_t/Q$.}
\label{fig:ncollinear_Wilsonline}
\end{figure}

Our formalism is based on a factorized approach for boosted top jet production which, here we discuss exemplarily in the context of $e^+ e^-$-collisions for the double hemisphere top jet mass distribution. In the limit where the top is produced with large center-of-mass energies $Q \gg m_t$ one can use effective field theory methods to disentangle the different scales in the production and decay process. One can show that, when the decay final state products are in the top resonance region (i.e.~$(p_b+p_W)^2-m_t^2\ll m_t^2$), the $W$-boson is emitted collinear to the top-direction and thus the diagrams of the off-shell computation involve eikonal propagators at leading order in $m_t/Q$ that arrange themselves into collinear Wilson lines (see Figure \ref{fig:ncollinear_Wilsonline}) that are independent of the top production mechanism. 

\begin{figure}[!h!tbp]
\centering
\subfigure[]{\includegraphics[height=5cm]{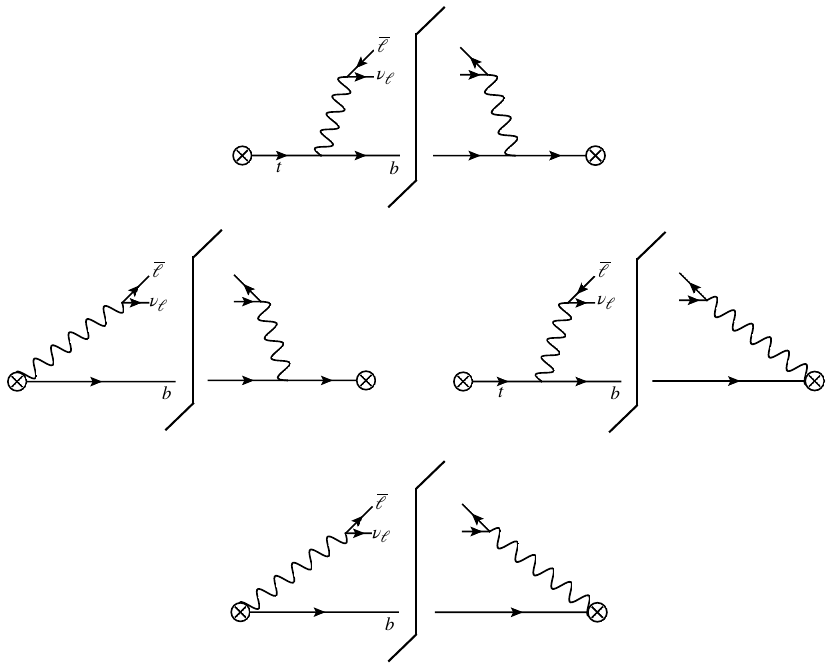} \label{fig:differentialjetfct_a}}\quad
\subfigure[]{\includegraphics[height=5cm]{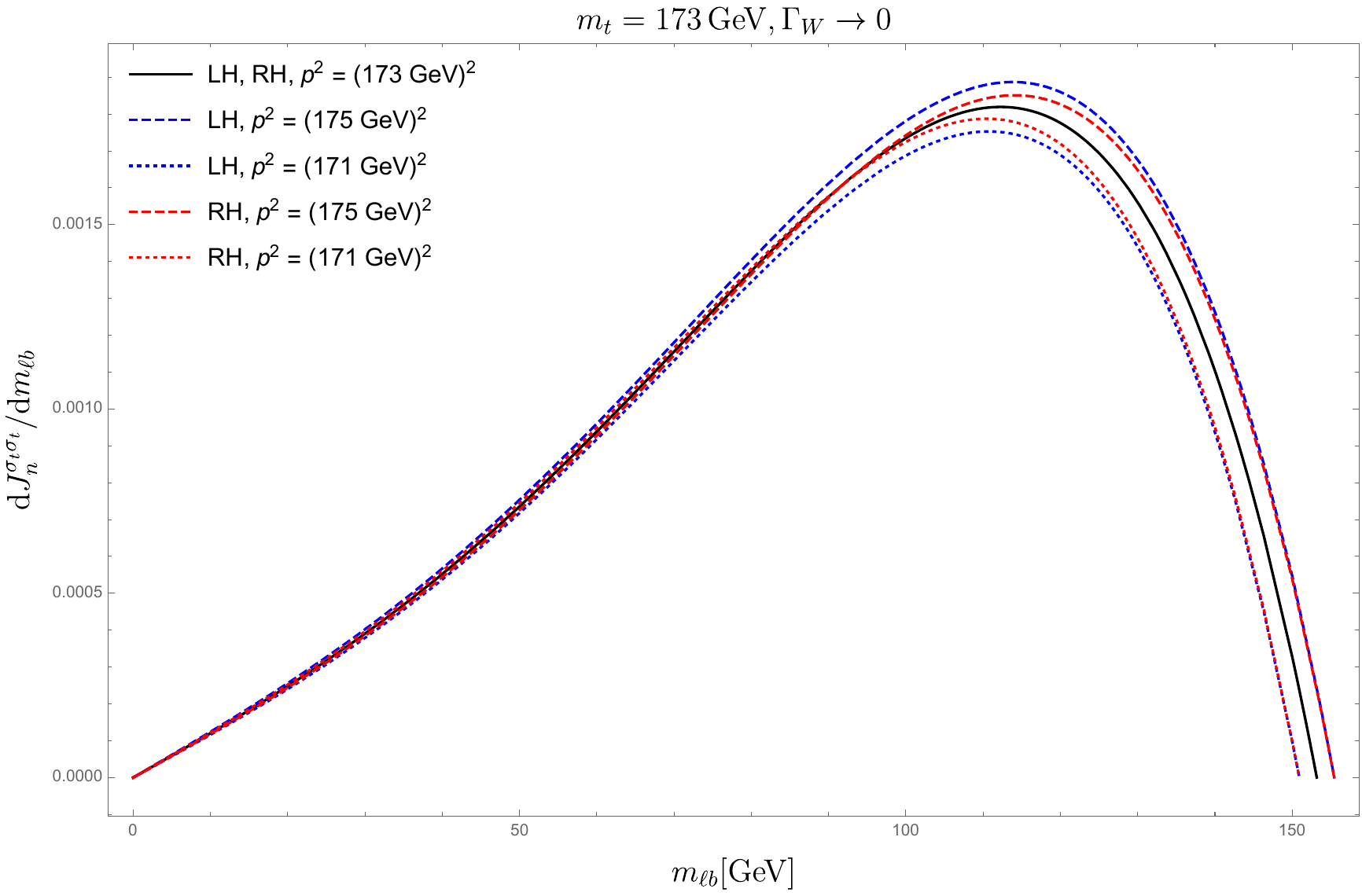} \label{fig:differentialjetfct_b}}
\caption{Feynman diagrams pertaining to the tree-level jet function in (a) and in (b) the differential distributions in $m_{\ell b}$ of the chirality conserving jet functions ($\sigma_t=\mathrm{RH}$ for right-chiral and $\sigma_t=\mathrm{LH}$ for left-chiral jet functions) for different values of the off-shellness of the top are shown.}
\label{fig:differentialjetfct}
\end{figure}

\begin{figure}[!h!tbp]
\centering
\includegraphics[height=7.5cm]{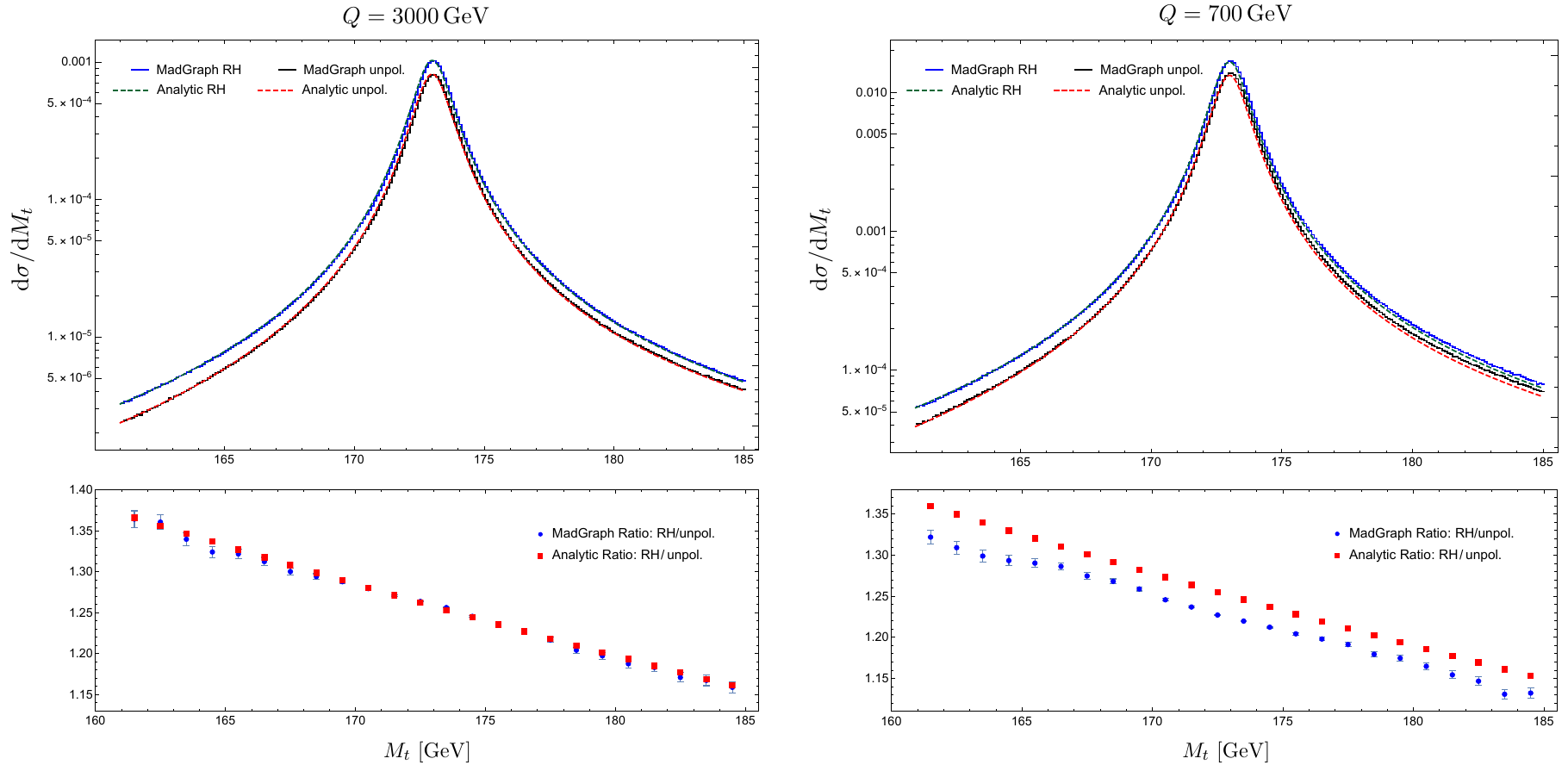}
\caption{Distributions of cross-section for $e^- e^+ \rightarrow b W^+ \overline{t}$ with respect to the $bW$-invariant mass $M_t$ for different CM energies and for different polarisations of the initial state as well as ratio-plots for these polarisations. We compare the results generated in \MadGraph~to the effective theory approach.}
\label{fig:MGcomparison}
\end{figure}

As a result, one can derive a factorization theorem for the top invariant mass spectrum in the resonance region $s_{t,\bar{t}}\equiv M_{t,\bar{t}}^2-m^2_t \ll m_t^2$ \cite{PhysRevD.77.074010, Dicus:1984fu}
\begin{equation}
\begin{split}
\frac{\mathrm{d}^2 \sigma}{\mathrm{d} M_{t}^2 \, \mathrm{d} M_{\bar{t}}^2 \mathrm{d}\Omega_t} \sim & \sum_{\sigma_t,\sigma_t'} \sum_{\sigma_{\overline{t}},\sigma_{\overline{t}}'} \mathcal{M}_{Q}(\sigma_t, \sigma_{\overline{t}}) \mathcal{M}^*_{Q}(\sigma_t',\sigma_{\overline{t}}')  \\ 
& \times \int \mathrm{d} l^+ \, \mathrm{d} l^- \, J_t^{\sigma_t \sigma_t'} (s_t - Q l^+) \, J_{\bar{t}}^{\sigma_{\overline{t}} \sigma_{\overline{t}}'} (s_{\bar{t}} - Q l^-) \, S_{\mathrm{hemi}} (l^+, l^-) \ ,
\end{split}
\end{equation}
where $M_t^2$ and $M_{\bar{t}}^2$ are the invariant hemisphere masses for the top and the antitop jets respectively and $\mathcal{M}_Q(\sigma_t, \sigma_{\overline{t}})$ denotes the matrix element for on-shell top pair-production at the hard scale $Q$ with the top (anti-top) helicity $\sigma_t (\sigma_{\overline{t}})$. The solid angle $\Omega_t$ is defined with respect to the $t\overline{t}$-axis. At tree-level the soft function is given by $S_\mathrm{hemi} (l^+, l^-) = \delta(l^+) \delta (l^-)$. 
\\
The central objects for the discussion of the top decay are the top and antitop jet functions, $J_t^{\sigma_t \sigma_t'}$ and $J_{\bar{t}}^{\sigma_{\overline{t}} \sigma_{\overline{t}}'}$. The chirality-conserving ($\sigma_t\!=\!\sigma_t'$) $n$-collinear top jet functions can be written as
\begin{equation}
J_t^{\sigma_t \sigma_t}(p^2)=\frac{\sum_X (2\pi)^3 \delta^{(4)}(p-P_X)}{N_C(\overline{n}\cdot p)}\mathrm{Tr}\left[\bra{0}\frac{\slashed{\overline{n}}}{4} \chi_n^{\mathrm{\sigma_t}}(0)\ket{X}\bra{X}\overline{\chi^{\sigma_t}_n}(0)\ket{0}\right]\ ,
\label{Eq:Jetfunction} 
\end{equation}
where the left-/right-chiral jet field is given by $\chi_n^{\mathrm{\sigma_t}}=\omega_{\mp} \chi_n$ with the chirality projectors $\omega_{\mp}$ and with the light-cone directions $n^\mu=(1, \vec{p}_t/|\vec{p}_t|)$ and  $\overline{n}^\mu=(1, -\vec{p}_t/|\vec{p}_t|)$  along the top momentum axis $\vec{p}_t/|\vec{p}_t|$. If we consider a top quark decaying into a bottom quark and lepton pairs and evaluate the jet-correlator using Wick contractions, we find a universal and gauge-invariant jet function which describes off-shell top quark production including spin-correlations. In Figure \ref{fig:differentialjetfct_a} the squared matrix elements arising from Eq.~(\ref{Eq:Jetfunction}), contributing to the jet function at tree-level are shown. Due to the inclusion of the Wilson line contributions we get a gauge-invariant result (also for $p^2=(p_b+p_W)^2\!\neq\! m_t^2$) and we take into account the dominant off-shell effects in an expansion in $m_t/Q$.  
 \\
In Figure \ref{fig:differentialjetfct_b} we plot the differential chirality conserving jet functions with respect to the $b$-jet lepton invariant mass $m_{\ell b}$ for different values of the off-shell top quark invariant mass $\sqrt{p^2}$ and a fixed top mass $m_{t}\!=\!173 \, \mathrm{GeV}$ in the NW limit of the $W$-boson (i.e. $\Gamma_W\rightarrow 0$). We observe that the endpoint of the distribution shifts with the off-shellness of the top as expected, but it is independent of the chirality. On the other hand, the maximum of the distribution is $p^2$- and chirality-dependent.  \\
A comparison of the cross-section for $e^- e^+ \rightarrow b W^+ \overline{t}$ in our approach to the full tree-level calculation as given by \MadGraph~\cite{Alwall_2014} shows very good agreement for highly boosted top quarks, i.e.~at high center-of-mass energies $Q \gg m_t$ such as $Q\!=\!3000$ GeV. For simplicity, we only treat the top as an unstable particle. 
As expected, for much lower values of $Q$ (e.g. $Q\!=\!700$ GeV), taking into account only leading power effects in the expansion in $m_t/Q$ leads to a difference between the two approaches. \\
This behaviour is shown in Figure~\ref{fig:MGcomparison}, where we have plotted the $M_t$ differential cross-section for an unpolarised initial state as well as with a right-handed polarised electron ($P_{e^-}\!=\!1$) and a left-handed positron ($P_{e^+}\!=\!-1$) which yields the right-chiral (RH) jet function. Taking the ratio of the cross-sections for the two polarisations of the initial state and comparing our analytic results to the ones of \MadGraph \,(see Figure~\ref{fig:MGcomparison}, below cross-section distributions), we find that the shape information of the distributions agree to very good precision and that subleading power corrections arising for $Q\!=\!700$ GeV mostly affect the normalisation of the prediction given by the factorization theorem.
\section{Conclusions and outlook}
In this presentation we discussed the basic concepts of our new formalism that allows to combine the factorization property of the NW limit and off-shell computations for boosted top quark production and its subsequent decay. Our approach allows to study top decay sensitive observables and, eventually, enables us to consistently incorporate resummed QCD corrections. For the QCD corrections we will use a subtraction method and perform the phase-space integration with a Monte-Carlo code. Further details about our formalism will be presented in~\cite{Hoang_2023}.\\

\Acknowledgements
This work was supported in part by the FWF Austrian Science Fund under the Project No. P32383-N27. We also acknowledge partial support by the FWF Austrian Science Fund under the Doctoral Program ``Particles and Interactions" No. W1252-N27 as well as the Erwin Schr{\"o}dinger International Institute for Mathematics and Physics. We thank the Faculty of Physics of the University of Vienna.

\bibliography{eprint}{}
\bibliographystyle{unsrt}
 
\end{document}